\documentstyle [11pt] {article}

\begin{document}

\begin{center}{\bf Electrostatic Energy Calculations}
\end{center}
\begin{center}{\bf for Molecular Dynamics}\end{center}
\vspace{0.35in}

Michael J. Love and Henri J.F. Jansen

Department of Physics, Oregon State University,
Corvallis OR 97331

\vspace{0.35in}

The evaluation of Coulomb forces is a difficult task. The
summations that are involved converge only conditionally
and care has to be taken in selecting the appropriate
procedure to define the limits. The Ewald method is a
standard method for obtaining Coulomb forces, but this
method is rather slow, since it depends on the square of
the number of atoms in a unit cell. In this paper we have
adapted
the plane-wise summation method for the evaluation of Coulomb
forces. The use of this method allows for larger computational
cells in molecular dynamics calculations.

\vspace{0.5in}


\newpage
\noindent{\bf I. Introduction}

     When periodic boundary conditions are employed,
a molecular dynamics (MD)
computational cell containing N atoms may be
considered to be a unit cell
with an N-point basis in a crystal lattice.
The calculations of long-range
Coulomb interactions can then be performed with
techniques developed for the
evaluation of electrostatic energies and forces
within crystals. While
essentially all molecular dynamics simulations
which include long-range
ionic interactions have used the Ewald method$^{1}$
to calculate the Coulomb
energy and forces, alternative methods offer some computational
advantages.  In the following sections, calculations
of lattice sums with
Ewald, planewise summation, and multipole techniques
are briefly described and compared.
The convergence properties of the Ewald summation
are shown to be due to effective surface charges
included implicitly in the sum
and a multipole formulation is presented which
produces results identical to
those obtained with the Ewald method.

\noindent{\bf II. Lattice summation methods}

In the Ewald method, the potential due to a
lattice of unit point
positive charges is obtained
by surrounding each point charge with
equal positive and negative Gaussian charge
distributions and a
uniform negative volume charge. For negative
charges one simply
reverses the signs of all charges.
 The point charges plus the
negative Gaussian distributions and the positive
Gaussian distributions
plus the negative volume charges are then summed
separately, with the
convergence of the positive Gaussian-volume charge
lattice sum improved
by performing the sum in reciprocal space.

When the energy of a collection of positive and
negative point charges in the unit cell of a neutral crystal
is calculated using the Ewald potential, the background
charges cancel
in the volume of the crystal
and the energy sum is generally taken to be that due to the
point charges alone.

The overall
speed of convergence for Ewald sums can be improved in a
simple way by adjusting the
free parameter in the equations so that the direct and
reciprocal
sums converge at the same rate.  Nijboer and de Wette$^{2}$
have shown
that this will occur if $\alpha=\sqrt{\pi}$/L, where L is
the length of
a cubic cell edge and $\alpha$ is the free parameter.
Sangster and Dixon$^{3}$ show that the reciprocal sum
can be reformulated so that the number of terms is
proportional to N
rather than N$^{2}$, and the free parameter can then be
adjusted to
increase the number of reciprocal space terms and decrease
the number of
direct space terms required to obtain a given degree of
convergence.
The optimum value of $\alpha$ will be that which minimizes
the overall
computational time.  An implementation of both the original
and modified
Ewald methods$^{4}$ indicates that the techniques suggested
by Sangster and
Dixon can increase the speed of computation by a factor of
about two or
three.

   A drawback of using the Ewald method for computing
Coulomb forces presents itself
when simulations containing large numbers of particles
are considered.
For an MD cell containing N atoms, the number of Ewald
sums required to
calculate the forces is proportional to $N^{2}$.
The terms which are summed in an Ewald energy
calculation are rather
complicated, particularly the direct sum.
A force calculation requires the
calculation and summation of terms which are
even more complicated,
and the repeated calculations of
lattice sums in a molecular dynamics calculation
can be very time consuming,

    An alternative method for performing Coulomb
lattice sums is the planewise
summation method (PSM)
first developed by Nijboer and de Wette$^{5}$
for dipole lattices and crystals with monoclinic
and higher symmetry.
In this method, lattice summations are computed by applying
a two-dimensional Fourier transform over two indices and then
performing the summation over the third index analytically.
There are no
additional charges introduced to speed the convergence.

The PSM was later extended to include summation of
multipole lattices of all
orders$^{6}$,
summation of multipole
lattices in triclinic crystals$^{7}$, and the direct
calculation of
the electrostatic potential$^{8}$.

     The planewise summation method offers some
computational advantages over
the Ewald method. The lattice summation is
performed over two indices
rather than three, there is only a reciprocal
space sum and not a direct
sum, and the individual terms in the sum are
generally simpler to calculate
than those in an Ewald sum.

   The number of calculations required to find the
electrostatic energy by
direct planewise summation is proportional to N$^{2}$.
An additional
disadvantage of the direct planewise method is the
conditional convergence of the
lattice sums, leading to a dependence of the total energy
on the choice of the planes in the crystal and
inconsistency with periodic
boundary conditions when the dipole moment of the
unit cell is nonzero.

   The conditional convergence of the PSM is due to the
general convergence properties
of the Coulomb lattice sum for a neutral unit cell.
The energy sum may be written:

\begin{equation}
   E_{coul}=\frac{e^{2}}{2}
\sum_{\alpha=-\infty}^{\infty}\sum_{\beta=-\infty}^{\infty}
    \sum_{\gamma=-\infty}^{\infty}
    \sum_{i=1}^{N}\sum_{j=1}^{N}{}^{^{\prime}}
\frac{q_{i}q_{j}}{\mid \vec{r}
    _{i}-\vec{R}_{\alpha\beta\gamma}-\vec{r}_{j}\mid}
\end{equation}

\noindent where
$\vec{R}_{\alpha\beta\gamma}
\equiv\alpha\vec{a}+\beta\vec{b}+\gamma\vec{c}$;
$\vec{a}$, $\vec{b}$, and $\vec{c}$ are Bravais
lattice vectors; $q_{i}$ is the valence
of the i$^{th}$ atom; and the prime on the
summation over lattice indices indicates
omission of the terms for which
$i=j$ and $\alpha = \beta = \gamma =0 $.

    Using this expression to approximate the energy
by evaluation of a finite number
of terms leads to contradictions with the
periodic boundary conditions which
are commonly imposed on a MD cell. In the
usual case of a cell in which the total charge
is zero, the energy sum will approach a finite limit.
However,
if the cell has a nonzero dipole
moment there will be a constant electric
field component throughout the
MD cell, so that the potential at a point
on one face of the cell
will differ from that at a point translated
through the cell along a Bravais
lattice vector to the opposite face.
The existence of a non-zero dipole moment also makes
the constant electric field component
dependent on the choice of the
unit cell.  An example of how this can
occur is illustrated in Figure 1.
Two different, equivalent choices for the unit
cell of a tetragonal AB$_{2}$ ionic
compound are shown along with the net dipole moment
for the unit cell.  It is seen that the dipole moment
changes sign if
the locations of the atoms on the corners of the cell
are redefined.
If the derivative of the energy sum is calculated for
each of these cells,
then each sum will contain terms describing an electric
field parallel to the
dipole moment of the associated unit cell, and even
though the cells are
physically equivalent the electric fields for the two
cases will differ
in sign.

In an MD simulation with periodic boundary conditions,
if a particle leaves the unit cell
then there is an identical particle
which simultaneously enters the cell through
the opposite face. If the exiting particle is replaced
in the
energy sum by the entering particle, the unit cell
is effectively
redefined. The configuration of the unit cell could change,
for example,
from one of those shown in Figure 1 to the other
in the course of the
simulation.  When this happens, the electric field inside the
cell changes discontinuously.
 Since the particles in an MD cell are not
taken to have any particular symmetry,
a dipole moment is generally
present and discontinuities in the electric
field and therefore the forces
is the general rule.

   The dependence of the energy sum on the dipole
moment of the unit cell
is well known.  In addition to early studies on
the effect of the order of
summation of terms on the energy totals$^{9}$,
the development of fast
summation techniques for the Coulomb energy of ionic
lattices has led to
methods which can give different total energy values
for the same unit cell.
The Ewald method and the planewise summation method, for
example, yield different values if the unit cell has
a net dipole moment.

    An alternative to both the Ewald and the planewise method
as discussed above begins with the separation of the Coulomb
potential into multipoles. The resulting multipole
lattice sums can then be summed
by the planewise method. The lattice sums do not
contain the atomic
positions, so if simulations are carried out in a
unit cell with fixed shape,
these sums are constant and need to be calculated only once.
For simulations in MD cells with variable shape,
the lattice sums must be
recalculated at each time step.  In either case, using
procedures similar to those of the fast multipole
method of Greengard
and Rokhlin$^{10}$ it is possible to construct
algorithms for the forces which require
a number of calculations proportional to N.

   Although a multipole expansion of the Coulomb
energy contains the same
indeterminacy and dipole dependence of the direct
Coulomb sum, a simple modification
can be made which results in lattice sums which are all
absolutely convergent and which gives results
identical to those obtained by the
Ewald method.  In the next section, the
Ewald energy sum is derived  as the limit of
a finite lattice summation, and the magnitude of
the implicit surface charges is
found.  The following section describes the modified
multipole formulation and
shows the equivalence of this method to the Ewald method.

\noindent {\bf III. The Ewald Method for Finite Lattices}

   Since Ewald lattice sums are absolutely
convergent while direct Coulomb
lattice sums can be conditionally convergent,
there must be contributions
to the Ewald sums in addition to those from the
original point charges and
their images. By replacing the infinite
lattice summation in the Ewald method with a
summation over a very large
but finite lattice, the additional terms can be
expressed as surface charges
located on the faces of the volume occupied by the lattice.

   The Ewald potential includes a sum of terms from
point charges, positive and negative
Gaussian charge distributions, and a uniform background
charge.  The Gaussian
distributions cancel at each lattice point, so
the electrostatic potential at a point $\vec{r}$ is
due only to a lattice of unit point
charges plus a neutralizing background charge.
This potential is:

\begin{equation}
 \Phi_{ew}(\vec{r}) \equiv
\lim_{M\rightarrow\infty}\sum_{\alpha\beta\gamma =-M}^{M}
       \left\{ \frac{1}{\mid
\vec{R}_{\alpha\beta\gamma}-\vec{r}\mid}
       -\frac{1}{V}\int_{V_{\alpha\beta\gamma}} d^{3}r^{\prime}
        \frac{1}{\mid
\vec{R}_{\alpha\beta\gamma}+\vec{r}^{\prime}-\vec{r}\mid}
\right\}
\end{equation}

\noindent $V$ is the volume of the unit cell
and $V_{\alpha\beta\gamma}$ is the volume around
the lattice point $\vec{R}_{\alpha\beta
\gamma}$. In order to simplify the expression for
the potential, the limits of the
summations are taken to be the same for all three directions.
The lattice sum converges with increasing M, although
the value of the sum
 depends
on the manner in which the limits are approached and
will be different if
we take the limits $M_\alpha, M_\beta, and M_\gamma$
in a different way.

When the energy of a collection of positive and
negative point charges in the unit cell of a neutral crystal
is calculated using the Ewald potential, the background
charges cancel
in the volume of the crystal.
There are, however, regions on the surfaces of a finite
lattice in which the
overlapping volume charges do not cancel.
The magnitude of these charges is found
by consideration of the volume charge contribution
to the electrostatic energy sum.

     The electrostatic energy in the Ewald formulation
is equal to
\begin{eqnarray}
  E_{ew} & = & \frac{e^{2}}{2}\sum_{i=1}^{N}
\sum_{j=1}^{N}\Phi_{ew}(\vec{r}_{i}-\vec{r}_{j})  \nonumber \\
         & \equiv & E_{coul}+E_{surf}^{(ew)}  \nonumber \\
  E_{coul} & = & \frac{e^{2}}{2}
\lim_{M\rightarrow\infty}\sum_{\alpha\beta\gamma =-M}^{M}
       \sum_{i=1}^{N}\sum_{j=1}^{N}{}^{^{\prime}} q_{i}q_{j}
             \frac{1}{\mid
\vec{R}_{\alpha\beta\gamma}-\vec{r}_{j}+\vec{r}_{i}\mid} \\
  E_{surf}^{(ew)} & = & -
\frac{e^{2}}{2}\lim_{M\rightarrow\infty}
\sum_{\alpha\beta\gamma =-M}^{M}
       \sum_{i=1}^{N}\sum_{j=1}^{N} q_{i}q_{j}
       \frac{1}{V}\int_{V_{\alpha\beta\gamma}} d^{3}r^{\prime}
        \frac{1}{\mid
\vec{R}_{\alpha\beta\gamma}+\vec{r}^{\prime}-
\vec{r}_{j}+\vec{r}_{i}\mid}
     \nonumber
\end{eqnarray}

   The energy due to the uniform volume charge
integration can be more simply expressed
by making a change of variables:

\begin{eqnarray}
 \vec{r}^{\prime\prime} & = &
\vec{r}^{\prime}+\vec{r}_{i}+
\vec{R}_{\alpha\beta\gamma} \nonumber \\
       & = & \xi^{\prime\prime}\vec{a}+
\eta^{\prime\prime}\vec{b}+
       \zeta^{\prime\prime}\vec{c} \nonumber \\
       \vec{r}_{i} & = &
\xi_{i}\vec{a}+\eta_{i}\vec{b}+\zeta_{i}\vec{c}  \\
       \vec{r}_{j} & = &
\xi_{j}\vec{a}+\eta_{j}\vec{b}+\zeta_{j}\vec{c} \nonumber
\end{eqnarray}

\noindent  $\xi$, $\eta$, and $\zeta$
are the displacements along the $\vec{a}$, $\vec{b}$, and
$\vec{c}$ directions, respectively.

\noindent With these substitutions,

\begin{equation}
   E_{surf}^{(ew)}=
\lim_{M\rightarrow\infty}
\frac{-e^{2}}{2}\left\{ \sum_{j}q_{j}\sum_{i}q_{i}
       \int_{-M-\frac{1}{2}+\xi_{i}}^{M+
\frac{1}{2}+\xi{i}}d\xi^{\prime\prime}
       \int_{-M-\frac{1}{2}+\eta_{i}}^{M+
\frac{1}{2}+\eta{i}}d\eta^{\prime\prime}
       \int_{-M-\frac{1}{2}+\zeta_{i}}^{M+
\frac{1}{2}+\zeta{i}}d\zeta^{\prime\prime}
       \frac{1}{\mid\vec{r}^{\prime\prime}-
\vec{r}_{j}\mid} \right\}
\end{equation}

With a neutral unit cell, the volume charge is
zero within the limits $-A$
to $+A$, $-B$ to $+B$, and $-C$ to
$+C$.  If this volume is subtracted from the sum,
the contributions from the
volume charges are limited to integrations over cells
which are at the limits of the
lattice sums.  If the limits are sufficiently large,
the volume charge may be
approximated by a surface charge concentrated at the limit.
For example,
with M sufficiently large,

\begin{eqnarray}
 -\sum_{i}\int_{M}^{M+\frac{1}{2}+\xi_{i}}
\frac{d\xi^{\prime\prime}}
         {\mid\vec{r}^{\prime\prime}-
\vec{r}_{j}\mid} & \approx &
      -\sum_{i}q_{i}\frac{M+\frac{1}{2}+\xi_{i}-M}
          {\mid\vec{r}^{\prime\prime}_{
\eta\zeta}+M\vec{a}-\vec{r}_{j}\mid}  \nonumber \\
     & = & \frac{-1}{\mid\vec{r}^{
\prime\prime}_{\eta\zeta}+M\vec{a}-\vec{r}_{j}\mid}
           \sum_{i}q_{i}\xi_{i}
\end{eqnarray}

\[     \vec{r}^{\prime\prime}_{\eta\zeta}
\equiv \eta^{\prime\prime}\vec{b}+
                    \zeta^{\prime\prime}\vec{c}    \]

\noindent The potential due to the surface charge
is found by performing the
surface integration over $\eta^{\prime\prime}$
and $\zeta^{\prime\prime}$.
Neglecting edge and corner effects, the total
energy due to the surface charges is

\begin{eqnarray}
  E_{surf}^{(ew)} & = & \lim_{M\rightarrow\infty}
\sum_{j}q_{j}\left\{
     \int_{-M}^{M}d\eta\int_{-M}^{M}d\zeta \left[
        \frac{\rho_{a}}{\mid\vec{r}_
{\eta\zeta}+M\vec{a}-\vec{r}_{j}\mid}
       -\frac{\rho_{a}}{\mid\vec{r}_
{\eta\zeta}-M\vec{a}-\vec{r}_{j}\mid} \right] \right.
\nonumber \\
   & &  +\int_{-M}^{M}d\xi\int_{-M}^{M}d\zeta \left[
        \frac{\rho_{b}}{\mid\vec{r}_
{\xi\zeta}+M\vec{b}-\vec{r}_{j}\mid}
       -\frac{\rho_{b}}{\mid\vec{r}_
{\xi\zeta}-M\vec{b}-\vec{r}_{j}\mid} \right] \nonumber \\
   & &  \left. +\int_{-M}^{M}d\xi\int_{-M}^{M}d\eta \left[
        \frac{\rho_{c}}{\mid\vec{r}_
{\xi\eta}+M\vec{c}-\vec{r}_{j}\mid}
       -\frac{\rho_{c}}{\mid\vec{r}_
{\xi\eta}-M\vec{c}-\vec{r}_{j}\mid} \right]
        \right\}
\end{eqnarray}

\[ \rho_{a}= -\sum_{i}q_{i}\xi_{i}   \]
\begin{equation}
   \rho_{b}= -\sum_{i}q_{i}\eta_{i}
\end{equation}
\[ \rho_{c}= -\sum_{i}q_{i}\zeta_{i}   \]

   If the surface charge is approximated by
a point charge of the appropriate
value at each surface lattice point, the
integrals can be replaced with two-dimensional
lattice sums.  The value of the point charge is
found to be numerically equal to that
of the surface charge.  As the limits are taken to
infinity, the volume, surface and
point charge representations become equivalent.

By inspecting the surface of the finite crystal,
it can be seen that the volume charges cancel each
other exactly only if
the net dipole moment of the unit cell is zero.
If the dipole moment is
nonzero, then  an Ewald summation contains
contributions from surface charges
with magnitudes given by equation (8) in addition
to the direct Coulomb sum of equation (1).

\noindent {\bf IV. Multipole Lattice Sums}
\normalsize

   In terms of the unnormalized spherical harmonics,
the inverse distance between
two points $\vec{r}$ and $\vec{r}^{\prime} +
\vec{R}_{\alpha\beta\gamma}$ is$^{6}$:

\begin{equation}
    \frac{1}{\mid \vec{r}
    -\vec{R}_{\alpha\beta\gamma}-\vec{r}^{\prime}\mid} =
   \sum_{k=0}^{\infty}\sum_{n=-k}^{k}
   \sum_{l=0}^{\infty}\sum_{m=-l}^{l}
   \frac{(-1)^{l}(k+l-m-n)!}{(k+n)!(l+m)!}
     r^{l}Y_{lm}^{*}(\hat{r})
   r^{\prime k}Y_{kn}^{*}
(\hat{r}^{\prime})\frac{Y_{k+l,n+m}
(\hat{R}_{\alpha\beta\gamma})}
   {R_{\alpha\beta\gamma}^{k+l+1}}
\end{equation}

\begin{equation}
Y_{lm}(\hat{r})\equiv P_{l}^{\mid m \mid}(cos \theta)e^{im\phi}
\end{equation}

  This expression is only valid if $r+r^{\prime}
< R_{\alpha\beta\gamma}$.
   Because of this, it is
necessary to define a sphere about the origin of the
MD cell and restrict
the application of the multipole expansion to those terms
in the lattice
sums which come from cells with centers outside the sphere.
This radius must be greater than the longest diagonal of
the MD cell
to insure the validity of the multipole expansion in all
 cases. Contributions
from `near neighbors' cells inside the sphere are summed
directly, so that the
total Coulomb energy sum is:

\begin{eqnarray}
 E_{coul} & = & \frac{e^{2}}{2}\sum_{\alpha\beta\gamma=nn}
    \sum_{ij=1}^{N}{}^{^{\prime}}\frac{q_{i}q_{j}}{\mid \vec{r}
    _{i}-\vec{R}_{\alpha\beta\gamma}-\vec{r}_{j}\mid} \\
 & & +\frac{e^{2}}{2}\sum_{\alpha\beta\gamma\not=nn}
    \sum_{i=1}^{N}\sum_{j=1}^{N}q_{i}q_{j}
   \sum_{k}\sum_{n}
   \sum_{l}\sum_{m}
  c_{kn;lm}r_{i}^{l}Y_{lm}^{*}(\hat{r}_{i})
   r_{j}^{k}Y_{kn}^{*}(\hat{r}_{j})\frac{Y_{k+l,n+m}
(\hat{R}_{\alpha\beta\gamma})}
   {R_{\alpha\beta\gamma}^{k+l+1}}    \nonumber \\
c_{kn;lm} & \equiv & \frac{(-1)^{l}(k+l-m-n)!}{(k+n)!(l+m)!}
 \end{eqnarray}

\noindent With the multipole moments of the MD cell
defined as

\begin{equation}
Q_{lm}\equiv\sum_{i=1}^{N}q_{i}
r_{i}^{l}Y_{lm}(\hat{r}_{i})
\end{equation}
and the first term in equation (11) defined as $E_{nn}$,
the energy is:

\begin{equation}
 E_{coul}=E_{nn}+\frac{e^{2}}{2}
    \sum_{kn}\sum_{lm}
   c_{kn;lm}Q_{lm}^{*}Q_{kn}^{*}
      \sum_{\alpha\beta\gamma\not=nn}
   \frac{Y_{k+l,n+m}(\hat{R}_{\alpha\beta\gamma})}
   {R_{\alpha\beta\gamma}^{k+l+1}}
\end{equation}

   The lattice sums can be calculated separately
for each combination of $k+l$
and $n+m$.  For $k+l\geq 4$, these sums are
absolutely convergent$^{11}$ and can be
calculated by the PSM without ambiguity.  The
indeterminancy in the total
energy is contained in the lattice sums with $k+l<4$,
and since these sums
are at best conditionally convergent other methods must
be used for their evaluation.

   In all cases but one, terms in the sum  which contain
combinations of $k$ and $l$ with $k+l<4$ can be shown to
be identically zero.
If either $k$ or $l$ is equal to zero, then the charge
neutrality of the MD cell insures that $Q_{00}=0$.
The inversion symmetry
of the Bravais lattice  in combination with the parity
of spherical
harmonics of odd order leads to zero contributions from all
odd-valued combinations, including $k+l$ equal to 1 or 3.
The only
non-zero terms with $k+l<4$ are those  with $k=1$ and $l=1$.
The strength of this term is determined by the
collective dipole
moment of the charges in the MD cell.  As expected,
if the dipole moment is zero,
then $Q_{1m}$ is zero and there is no indeterminancy
is the overall sum.

   In order to evaluate the term with $k=l=1$ and
nonzero dipole moment,
a set of point charges of arbitrary magnitude are
added and subtracted at the center of each face of
each cell in the
lattice.
The energy in the MD cell due to these added point charges is
zero, which in this case is written

\begin{equation}
 0=\frac{e^{2}}{2}\sum_{\alpha\beta\gamma=-M}^{M}
\sum_{i=1}^{N}q_{i}\sum_{j=a,b,c}
\{\frac{\tilde{q}_{j}}{\mid\vec{r}_{i}-
\vec{R}_{\alpha\beta\gamma}-\vec{\frac{j}{2}}\mid}
  -\frac{\tilde{q}_{j}}{\mid\vec{r}_{i}-
\vec{R}_{\alpha\beta\gamma}-\vec{\frac{j}{2}}\mid}\}
\end{equation}

\noindent or

\begin{eqnarray}
 0 & = & \frac{e^{2}}{2}\sum_{\alpha\beta\gamma=-M}^{M}
\sum_{i=1}^{N}q_{i}\sum_{j=a,b,c}
  \frac{\tilde{q}_{j}}{\mid\vec{r}_{i}-
\vec{R}_{\alpha\beta\gamma}-\vec{\frac{j}{2}}\mid} \nonumber \\
 & &  -\frac{e^{2}}{2}\sum_{\alpha=-M+1}^{M+1}\sum_
{\beta\gamma=-M}^{M}
                           \sum_{i=1}^{N}q_{i}
   \frac{\tilde{q}_{a}}{\mid\vec{r}_{i}-\vec{R}_
{\alpha\beta\gamma}+\vec{\frac{a}{2}}\mid} \nonumber \\
 & & -\frac{e^{2}}{2}\sum_{\beta=-M+1}^{M+1}\sum_
{\alpha\gamma=-M}^{M}
                           \sum_{i=1}^{N}q_{i}
   \frac{\tilde{q}_{b}}{\mid\vec{r}_{i}-\vec{R}_
{\alpha\beta\gamma}+\vec{\frac{b}{2}}\mid} \nonumber \\
& & -\frac{e^{2}}{2}\sum_{\gamma=-M+1}^{M+1}
\sum_{\alpha\gamma=-M}^{M}
                           \sum_{i=1}^{N}q_{i}
   \frac{\tilde{q}_{c}}{\mid\vec{r}_{i}-
\vec{R}_{\alpha\beta\gamma}+\vec{\frac{c}{2}}\mid}
\end{eqnarray}

    By combining the contributions from the
 positive and negative charges at
each common lattice point, a single lattice
sum plus a number of surface
terms are created:

\begin{eqnarray}
 0 & = & \frac{e^{2}}{2}
\sum_{\alpha\beta\gamma=-M}^{M}\sum_{i=1}^{N}
q_{i}\sum_{j=a,b,c}
\{\frac{\tilde{q}_{j}}{\mid\vec{r}_{i}-
\vec{R}_{\alpha\beta\gamma}-\vec{\frac{j}{2}}\mid}
  -\frac{\tilde{q}_{j}}{\mid\vec{r}_{i}-
\vec{R}_{\alpha\beta\gamma}+\vec{\frac{j}{2}}\mid}\}
   \nonumber \\
 & &  -\frac{e^{2}}{2}
\sum_{i=1}^{N}q_{i}
\sum_{\beta\gamma=-M}^{M}\{\frac{\tilde{q}_{a}}
   {\mid\vec{r}_{i}-\vec{R}_
{-M-1 \beta\gamma}+\vec{\frac{a}{2}}\mid} -\frac{\tilde{q}_{a}}
   {\mid\vec{r}_{i}-\vec{R}_
{M \beta\gamma}+\vec{\frac{a}{2}}\mid} \} \nonumber \\
 & &  -\frac{e^{2}}{2}\sum_
{i=1}^{N}q_{i}\sum_{\alpha\gamma=-M}^{M}\{\frac{\tilde{q}_{b}}
   {\mid\vec{r}_{i}-\vec{R}_
{\alpha -M-1\gamma}+\vec{
\frac{b}{2}}\mid} -\frac{\tilde{q}_{b}}
   {\mid\vec{r}_{i}-\vec{R}_
{\alpha M\gamma}+\vec{\frac{b}{2}}\mid} \}  \nonumber \\
& & -\frac{e^{2}}{2}\sum_{i=1}^{N}
q_{i}\sum_{\alpha\beta=-M}^{M}\{\frac{\tilde{q}_{c}}
   {\mid\vec{r}_{i}-\vec{R}_{
\alpha\beta -M-1}+\vec{\frac{c}{2}}\mid} -\frac{\tilde{q}_{c}}
   {\mid\vec{r}_{i}-\vec{R}_{
\alpha\beta M}+\vec{\frac{c}{2}}\mid} \}
\end{eqnarray}

   The last three terms contain point
charges spaced uniformly on the surfaces of the
finite crystal.  Representing these
terms by $-E_{surf}^{mp}$ and performing a multipole
expansion on the terms in the lattice
which are not near neighbors,

\begin{eqnarray}
  0 & = & \frac{e^{2}}{2}\sum_
{\alpha\beta\gamma=nn}\sum_{i=1}^{N}q_{i}\sum_{j=a,b,c}
\{\frac{\tilde{q}_{j}}{\mid
\vec{r}_{i}-\vec{R}_{\alpha\beta\gamma}-
\vec{\frac{j}{2}}\mid}
  -\frac{\tilde{q}_{j}}{\mid\vec{r}_{i}-
\vec{R}_{\alpha\beta\gamma}+\vec{\frac{j}{2}}\mid}\}
  -E_{surf}^{mp} \nonumber \\
 & & +\frac{e^{2}}{2}\sum_{kn}\sum_{lm}c_{kn;lm}Q_{lm}^{*}
\{\sum_{j=a,b,c}\tilde{q}_{j}(\frac{r_{j}}{2})^{k}
   (Y_{kn}^{*}(\hat{j})-Y_{kn}^{*}(-\hat{j}))\} \nonumber \\
 & &  \times\sum_{\alpha\beta\gamma\not=nn}
    \frac{Y_{k+l,n+m}(\hat{R}_{\alpha\beta\gamma})}
   {R_{\alpha\beta\gamma}^{k+l+1}}
\end{eqnarray}

   Referring to the first term in this equation
as $\tilde{E}_{nn}$ and the expression in brackets in
the multipole term as $\tilde{Q}_{kn}^{*}$,
this equation becomes:

\begin{equation}
 0=\tilde{E}_{nn} - E_{surf}^{mp} +
   \frac{e^{2}}{2}\sum_{kn}\sum_{lm}c_{kn;lm}Q_{lm}^{*}
\tilde{Q}_{kn}^{*}
   \sum_{\alpha\beta\gamma\not=nn} \frac{Y_{k+l,n+m}
(\hat{R}_{\alpha\beta\gamma})}
   {R_{\alpha\beta\gamma}^{k+l+1}}
\end{equation}

   Values for $\tilde{q}_{a}$, $\tilde{q}_{b}$,
and $\tilde{q}_{c}$ are determined
by setting $\tilde{Q}_{1m}=-Q_{1m}$.The resulting
values for the compensating charges are

\begin{eqnarray}
  \tilde{q}_{a} & = & -\sum_{i=1}^{N} q_{i}\xi_{i} \nonumber \\
  \tilde{q}_{b} & = & -\sum_{i=1}^{N} q_{i}\eta_{i} \\
  \tilde{q}_{c} & = & -\sum_{i=1}^{N} q_{i}\zeta_{i} \nonumber
\end{eqnarray}

\noindent Compensating charges for the unit cells of
 Figure 1 are
shown in Figure 2.

  In an orthorhombic lattice the compensating
charges are proportional to the dipole
moments of the unit cell.  In triclinic lattices
the values are proportional to the
projection of the total dipole moment along the
lattice vectors. Using these values
for the compensating charges, the dipole moment
 for the MD cell can be equated to
an expression containing terms which represent a
surface charge, a near neighbor sum,
and lattice sums which are all absolutely convergent:

\begin{eqnarray}
  & & \frac{1}{2}\sum_{m=-1}^{1}
\sum_{n=-1}^{1}c_{1n;1m}Q_{1m}^{*}Q_{1n}^{*}
   \sum_{\alpha\beta\gamma\not=nn}
 \frac{Y_{2,n+m}(\hat{R}_{\alpha\beta\gamma})}
   {R_{\alpha\beta\gamma}^{k+l+1}} \nonumber \\
  & = & \tilde{E}_{nn}+
   \frac{1}{2}\sum_{lm}
\sum_{kn}c_{kn;lm}Q_{lm}^{*}\tilde{Q}_{kn}^{*}
   \sum_{\alpha\beta\gamma\not=nn}
 \frac{Y_{k+l,n+m}(\hat{R}_{\alpha\beta\gamma})}
   {R_{\alpha\beta\gamma}^{k+l+1}}
   -E_{surf}^{mp}
\end{eqnarray}

   Here the sums on the RHS are restricted
to values of $k$ and $l$ for which
$k+l\geq4$. When this expression is inserted
 into the original Coulomb sum,
the required form for the total electrostatic
 energy is obtained.

\begin{eqnarray}
   E_{coul} & = & E_{nn}+\tilde{E}_{nn}-E_{surf}^{mp}
 \nonumber \\
   & &  +\frac{1}{2}\sum_{lm}\sum_{kn}c_{kn;lm}Q_{lm}^{*}
   (Q_{kn}^{*}+\tilde{Q}_{kn}^{*})
   \sum_{\alpha\beta\gamma\not=nn} \frac{Y_{k+l,n+m}
(\hat{R}_{\alpha\beta\gamma})}
   {R_{\alpha\beta\gamma}^{k+l+1}}
\end{eqnarray}

  The distance from the original MD cell to the
 surface charges can be taken to be
very large compared to the cell dimensions.
In this limit the point charges at
the summation limits can be well approximated as
 surface charges with values
$\rho_{a}=\tilde{q}_{a}/A_{bc}$,
$\rho_{b}=\tilde{q}_{b}/A_{ac}$ and
$\rho_{c}=\tilde{q}_{c}/A_{ab}$.
$A_{ab}$ is the area of the MD cell
face defined by lattice vectors $\vec{a}$ and
$\vec{b}$. $A_{ac}$ and $A_{bc}$ are defined similarly.

   If the energy of the system is taken as

\begin{equation}
    E_{mp}\equiv E_{coul}+ E_{surf}^{mp}
\end{equation}

\noindent then all of the lattice sums are
absolutely convergent and
the forces generated by $E_{mp}$ are periodic with
the lattice. Because of this, the value
 of the energy is not affected by the definition
of the unit cell.  In fact, $E_{mp}$ is the same as the
Ewald energy if the same lattice limits are
 used in both cases, and
the surface charge distributions
generated by each method are identical.

\noindent {\bf V. Calculations}
\normalsize

Analytically, the multipole and Ewald methods
produce identical numerical results
when applied to a given MD cell.  The choice of which to
use in dynamical simulations may be made by
considering the speed and
accuracy of the algorithms available for each of these methods.

     The multipole expressions for energy and
 force were incorporated
in a rather simple way into energy and force
 subroutines for molecular dynamics simulations.
The near neighbor interactions are calculated
directly, and the planewise
summation method is used to calculate the lattice sums
for the long-range
Coulomb interactions.  The resulting algorithm
 includes a number of
calculations proportional to N$^{2}$ due to the
near neighbor terms.
The multipole calculations also contain a number of terms
proportional to N as well as a number of
calculations proportional to the fourth power
of the highest order multipole index included in the sums.

The multipole expansion is valid
for cells with centers more than the maximum MD
 cell diagonal from the the origin, but
if this distance is used as a cutoff radius the
 multipole terms will converge very
slowly.  For maximum efficiency, the number of
direct calculations within a given
cutoff radius must be weighed against the number
 of terms in the multipole
expansion which are required to achieve a given accuracy.

In addition to the error associated with truncation
of the multipole expansion,
additional numerical errors arise from the computation
of the lattice sums using the planewise summation.
Because this method uses Fourier
transforms to replace multipole lattice sums with
 more rapidly converging
series, it can be applied only to
sums over a complete lattice.  In order to use the
 planewise method as part
of a multipole method, the multipole sums are written:

\begin{equation}
 \sum_{\alpha\beta\gamma\not =nn}\frac{Y_{lm}
(\hat{R}_{\alpha\beta\gamma})}
   {R_{\alpha\beta\gamma}^{l+1}} =
   \sum_{\alpha\beta\gamma}{}^{^{\prime}}
\frac{Y_{lm}(\hat{R}_{\alpha\beta\gamma})}
   {R_{\alpha\beta\gamma}^{l+1}} -
   \sum_{nn}{}^{^{\prime}}\frac{Y_{lm}(
\hat{R}_{\alpha\beta\gamma})}
   {R_{\alpha\beta\gamma}^{l+1}}
\end{equation}

   The primes denotes sums over all
lattice points except the origin.  The
sum over the complete lattice can be
calculated using the planewise summation method for each value
of $l$ and $m$.  The terms in the sum
over near neighbors are then calculated
separately and subtracted from the planewise result.

  The subtraction of two lattice sums which
 are nearly equal in value tends
to exaggerate the numerical errors introduced
with the planewise summation
method.  A truncation error in the planewise
 sums will produce errors relative
to the lattice sum over all space, which is
usually several orders of
magnitude larger than a sum which is restricted
 to lattice points outside a
cutoff radius.

For purposes of
comparison, subroutines were also written for
 calculation of the Coulomb energy
and forces with the Ewald method.
    The free parameter in the Ewald summations was set to
$\alpha=\sqrt{\pi}/a$, $a$ being the lattice constant.
 This definition of
$\alpha$ is used for all calculations described
here, with the result that for
non-cubic cells the rate of convergence depends
on the ratios of the lattice constants.

   Error estimates for various sets of computational parameters
required to obtain a specified error tolerance in
the multipole and Ewald
methods are listed in Table 1. The lattice sums
 for the planewise and Ewald summations
include all terms with the absolute value of any
 summation index less than or equal
to M$_{PSM}$ and M$_{ew}$, respectively.  $\mu$ is
the maximum multipole order included
in the multipole sums.

     In order to check the relative efficiency of the
 two methods, a number of computer runs were
made with each force routine for the same unit cell.
 Each run included
81 force and 3 energy calculations.  The parameters
 used were M$_{ew}$=3,
$\mu$=14 and M$_{PSM}$=7 for an expected relative
error of 10$^{-6}$.
Results are shown in Figure 3. More details are
found in reference 12.

\noindent {\bf V. Discussion}

The multipole subroutine described here includes a number
 of calculations proportional
to N$^{2}$ because of the direct terms in the lattice summation.
These terms are simpler than those in the Ewald
summation, and total run time is less for systems containing
moderately large numbers of atoms.
There are many approaches to improving the
speed of the fast multipole algorithm but these are
 not pursued here.
As implemented, the multipole routine is faster
than the Ewald routine but
not overwhelmingly so.  The relative simplicity
 of the Ewald formulation makes
the Ewald method easier to program and debug.

   In addition to the Coulomb interaction for
ionic crystals, it often desirable in
simulations to calculate the energy and forces from
interactions which are proportional to any negative power of
the separation distance.
     The Ewald method was extended by
 Nijboer and deWette$^{2}$ to include
this type of lattice sum, and  Williams$^{13}$
later extended their methods
to allow multiple atoms in a unit cell.
These techniques give formulas
which can be used to perform fast summation
of potentials and forces
which would otherwise converge very slowly,
such as the $r^{-4}$
dipole-charge interaction commonly encountered
in ionic models and
the $r^{-6}$ Van der Waals or dispersion interaction.
The formulas
become progressively more complicated as the reciprocal
 power increases,
and it is common to use a cutoff radius to calculate
these potentials
rather than utilize the rapidly converging formulation.

     Incorporation of interaction potentials of the
form r$^{-n}$ into a multipole
method presents some difficulties.  The multipole
separation used for the
Coulomb potential cannot be used for the higher
power terms; since these terms
do not satisfy Laplace's equation they cannot be
constructed from linear
combinations of the solutions to that equation.
The multipole character of the
summations could be preserved by separating the
lattice sums and particle
coordinates through application of a three-dimensional
Taylor series expansion.
These lattice sums could in principle be evaluated with
a number of
operations proportional to N by planewise summation and then
differentiated to obtain the terms which would be
included in the multipole sums.

     An alternative approach for including the
higher-power terms in a multipole method is found
by considering these terms in light of their
physical origins.  The r$^{-4}$
term is generally considered to be a charge-dipole
interaction, while the r$^{-6}$
terms usually arise from an induced dipole-dipole
interaction.  If each atom
is assigned a polarizability, then the dipole moment
induced in each atom will
be proportional to the electric field at that atom.
The field at each atom is
routinely obtained as part of the force calculation.
 The charge-dipole interaction could
then be calculated directly for near neighbors and
through an additional
multipole sum for the long-range contributions.
The corrections to the electric
field at each atom would in turn cause corrections
in the atomic dipole moments,
and repetition of this procedure would be necessary until
a specified degree of self-consistency is obtained.  At
each step the only corrections are to the multipole
moments of the unit cell;
the lattice sums remain unchanged throughout the
self-consistency routine.

     The direct calculation of dipole interactions
through the multipole method
would have a number of advantages.  All of the
electrostatic forces due to dipole
interactions are included automatically;
the r$^{-4}$ and r$^{-6}$ potentials,
for instance, need not be considered separately.
Using fast multipole
techniques, it is possible to construct algorithms
 which include all of the
electrostatic forces to any required degree of accuracy
with a number of
calculations proportional to N.  The short-range forces
would be calculated
in the last step of the fast multipole procedure and
would also require a
number of calculations proportional to N. The method
could readily be extended
to include quadrupole and higher order interactions if required.

\vspace{0.25in}

{\bf Aknowledgment}

This work was supported by the U.S.
Department of Energy (DOE) under grant
DE-FG06-88ER45352.

\newpage

{\bf References}

1. See, for example, J.C. Slater, "Insulators,
Semiconductores, and Metals" (McGraw Hill, New York, 1967),
chapter 9.

2. B. R. A. Nijboer and F. W. de Wette,
Physica {\bf 23}, 309 (1957).

3. M. J. L. Sangster and M. Dixon,
Adv. Phys. {\bf 25}, 247 (1976).

4. N. Karasawa and W. A. Goddard,
J. Phys. Chem. {\bf 93}, 7320 (1989).

5. B. R. A. Nijboer and F. W. de Wette,
 Physica {\bf 24}, 422 (1958).

6. B. R. A. Nijboer and F. W. de Wette,
Physica {\bf 24}, 1105 (1958).

7. V. Massidda and J. A. Hernando,
Physica {\bf 101B}, 159 (1980).

8. V. Massida, Physica {\bf 95B}, 317 (1978).

9. E. Evjen, Phys. Rev. {\bf 39}, 675 (1932).

10. L. Greengard and V. Rokhlin,
J. Comp. Phys. {\bf 73}, 325 (1987).

11. C. A. Scholl, Proc. Phys. Soc. {\bf 87}, 897 (1966).

12. M. J. Love, PhD thesis, 1993, unpublished.

13. D. E. Williams, Acta Cryst. {\bf A 27}, 452 (1971).

\newpage

Table 1.

Parameters for error tolerances for multipole and Ewald methods.

\vspace{0.5in}

\begin{center}
\begin{tabular} {|c|c|c|c|} \hline
$\epsilon_{coul}$  &  n=1.5    & n=2    & Ewald \\
  &$\mu$  \ \  $M_{psm}$ & $\mu$\ \ $M_{psm}$&$M_{Ew}$\\ \hline
$10^{-4}$  &  16   \ \  6  &  8  \ \    4  &  2  \\
$10^{-5}$  &  20   \ \  8  &  12   \ \   6  &  3  \\
$10^{-6}$  &  $>$20  \ \   $>$8  &  14  \ \    7  &  3  \\
$10^{-7}$  &  $>$20  \ \    $>$8  &16  \ \  8  &  4  \\ \hline
\end{tabular}
\end{center}

\newpage

{\bf Figure Captions}

Figure 1.

Two equivalent unit cells for tetragonal $ZrO_2$.
Cross-hatched circles
are $Zr^{4+}$ ions and open circles are $O^{2-}$ ions.
 P is the
dipole moment of the unit cell.

\vspace{0.5in}

Figure 2.

Addition of compensating charges which result in zero
dipole moments
for the unit cells of Figure 1. Parallel hatched circles are
the compensating charges.

\vspace{0.5in}

Figure 3.

Computer run times for various numbers of atoms per
unit cell. Forces
are calculated by the Ewald method (circles) or by the
planewise
summation method (triangles).

\end{document}